\documentclass[preprints,article,accept,moreauthors]{Definitions/mdpi} 

\firstpage{1} 
\pubvolume{1}
\issuenum{1}
\articlenumber{0}
\pubyear{2026}
\copyrightyear{2026}
\datereceived{ } 
\daterevised{ }
\dateaccepted{ } 
\datepublished{ } 

\Title{Flexible Piezoresistive Yarn Pressure Sensor for Arterial Blood Pressure Waveform Measurement}

\Author{Rizal Maulana $^{1,}$*\orcidA{}, Ádám Rák $^{1}$\orcidB{}, Sándor Földi $^{1}$\orcidC{} and György Cserey $^{1}$\orcidD{}}

\AuthorNames{Firstname Lastname, Firstname Lastname and Firstname Lastname}

\address{$^{1}$ \quad Faculty of Information Technology and Bionics, Pázmány Péter Catholic University, Práter u. 50/A,
1083 Budapest, Hungary; rizal.maulana@itk.ppke.hu (R.M.); rak.adam@itk.ppke.hu (A.R.); foldi.sandor@itk.ppke.hu (S.F.); cserey.gyorgy@itk.ppke.hu (G.C.)}

\corres{Correspondence: rizal.maulana@itk.ppke.hu}

\abstract{The arterial blood pressure (ABP) waveform contains essential information concerning the hemodynamic status and its relation to the diagnosis of cardiovascular conditions. Therefore, the continuous monitoring of the ABP waveform has become important. A measurement system that offers a non-invasive approach and high accuracy in detecting morphological details of the ABP waveform is needed. In this study, we introduce a flexible piezoresistive yarn (FPY) pressure sensor for continuous non-invasive ABP waveform measurement. The sensor was experimentally validated using two scenarios: measurements on a hardware simulator and at an arterial site. The simulator measurement aimed to assess the sensor’s performance in detecting consistent waveforms, while the arterial site measurements evaluated the sensor’s performance in measuring ABP waveforms corresponding to the cardiovascular cycle. The simulator measurement results showed high repeatability, reproducibility, and accuracy relative to the reference signal at the upper limit of realistic force magnitude, as measured by the correlation metric: 0.9999 ± 0.0002, 0.9994 ± 0.0002, and 0.9731 ± 0.0027, respectively. The correlation value derived from measurements at the arterial site compared with measurements using a 3D force sensor as a reference sensor was 0.9880 ± 0.0035. Based on our results, the FPY pressure sensor can be used as an innovative solution for monitoring ABP waveforms and has further potential to be developed as a diagnostic device for evaluating cardiovascular conditions.}


\begin{document}

\section{Introduction}
Continuous monitoring of arterial blood pressure (ABP) waveforms plays an important role in the clinical diagnosis of cardiovascular health and problems. ABP waveforms contain extensive information, and changes in their morphology over time can be analyzed to assess the hemodynamic status \citep{ref-1}. Along with providing the primary information on blood pressure itself, ABP waveforms also represent arterial stiffness, which greatly depends on the condition of the organs within the cardiovascular system. Because the morphological details of the ABP waveform are important diagnostic information \citep{ref-2}, the development of highly accurate measurement systems is essential. Furthermore, given the vital nature of cardiovascular health, early diagnosis of potential problems is imperative to facilitate prompt, life-saving medical interventions. To this end, the implementation of wearable devices offers an effective, continuous monitoring solution to achieve this early diagnosis.

Continuous ABP waveform measurement can be performed using the arterial cannula method, the gold standard of ABP measurement. This method involves inserting a catheter into a blood vessel to measure the intra-arterial pressure directly \citep{ref-3}. The invasive approach used in this method becomes a major concern because it poses a risk of vascular trauma. Therefore, an unobtrusive and non-invasive solution with comparable accuracy is needed. Multiple solutions for continuous non-invasive ABP waveform measurement have been developed in recent decades, including using indirect approaches through pulse waveform measurement. The pulse waveform is a morphologically equivalent signal to the ABP waveform, despite lacking absolute blood pressure values \citep{ref-4}. Although the amplitude of the pulse waveform does not directly correspond to physiological values without prior calibration, hemodynamic status can still be monitored by analyzing its various temporal features.

Uchida et al. \citep{ref-5} developed a non-invasive, contactless camera-based solution for pulse waveform measurement. A high-speed target spectral camera was used to capture a video image of the palm and face for the acquisition of RGB wavelength data. Three wavelength frequencies were measured in the video image, and the signals were averaged to generate the pulse waveform for each region of interest (ROI). However, the recorded pulse waveform lacked significant morphological details. This solution is also restricted to controlled environments, where lighting and the subject’s distance from the camera are regulated to provide a stable video image. Additionally, complex computations and high-performance hardware are needed for data processing.

Another popular non-invasive solution is the application of photoplethysmography (PPG) sensors. PPG is an optical sensor consisting of a light source and a photodetector used to measure blood volume variations in peripheral arteries. Research related to ABP waveform measurement typically implements PPG based on the Peñáz principle or its modifications \citep{ref-6,ref-7,ref-8}. The Peñáz principle is a volume-clamp-based method utilizing an inflatable finger cuff integrated with PPG. PPG detects variations in blood volume in the finger artery, and the resulting waveform serves as closed-loop feedback for the system to adjust the cuff pressure to maintain a steady blood volume during measurement. This method enables continuous ABP to be acquired by continuously measuring the cuff pressure using a manometer. The accuracy of the measurement, using PPG as the main sensor, depends on various factors, including sensor-skin contact, skin color, body temperature, and light conditions. The working mechanism of this method, which periodically adjusts cuff pressure, may result in numbness and arterial congestion in the fingers.

A natural solution for detecting physical phenomena in the form of pressure is by utilizing pressure sensors. In the implementation of ABP waveform measurement, the primary challenge of using a pressure sensor relates to the low-magnitude characteristics of the signal. A sensor with high sensitivity is a must, and an optimal measurement method is required to address this issue. A promising measurement method for non-invasive ABP waveform measurement is applanation tonometry \citep{ref-9,ref-10}. This method operates by applying light pressure to flatten (applanate) the superficial artery, which is adequately supported by the bone underneath. This procedure has demonstrated effectiveness in gaining pressure at the arterial site, resulting in sufficient pressure for sensor detection. Additionally, applying pressure perpendicular to the artery eliminates tangential forces on the arterial wall, ensuring that the measured pressure corresponds to intra-arterial pressure \citep{ref-11}. The radial artery at the wrist is the preferred site for applanation tonometry measurements. The main concern related to this method is the precision of sensor placement at the specified measurement point.

A variety of applanation tonometry devices are commercially available, including SphygmoCor and Millar SPT-301. However, these devices remain relatively expensive and rely on a hand-held sensor that is manually controlled for the applanation process. The measurement results are highly dependent on the user’s expertise in maintaining the stability of the sensor placement and the applied applanation pressure. Samartkit et al. \citep{ref-12} propose a solution for automating the placement of pressure sensors in the applanation tonometry method. In their study, piezoelectric pressure sensors made of lead zirconate titanate (PZT) ceramic material were used and attached to the wrist with a strap. Tightening the strap with adequate pressure can provide stable applanation pressure due to sensor surface pressure on the radial artery site. However, the rigid characteristics of PZT ceramic material and its large dimensions result in discomfort for the user when applanation pressure is applied for prolonged measurement periods.

Recent advancements in materials science have enabled the fabrication of flexible pressure sensors in various forms, including flexible fabric \citep{ref-13,ref-14,ref-15,ref-16}, flexible yarn \citep{ref-17,ref-18,ref-19,ref-20}, and others. The flexibility of these sensors can enhance user comfort during ABP waveform measurements. Moreover, due to its conformability, the sensor can effectively conform to the wrist’s contours, hence optimizing the precision of its placement on the radial artery site. These characteristics need to be implemented with a proper sensor design to achieve optimal sensitivity. Zhang et al. \citep{ref-21} fabricated a melamine sponge-covered graphene piezoresistive flexible pressure sensor for measuring pulse waveforms in the radial artery. The system has high accuracy in detecting heart rate and the actual blood pressure value, which is determined from the crest and troughs of the waveform. Nonetheless, the sensor is not sensitive enough to capture each detail of the pulse wave. Zhong et al. \citep{ref-22} proposed another solution by developing a knotted piezoresistive fiber-based flexible sensor using carbon nanotube/polyurethane terephthalate conductive yarn as the base material. However, sensor sensitivity was also an issue in this study. Both studies used an identical measurement protocol, utilizing bandages to attach the flexible sensor to the arterial site. This approach, combined with the design of each sensor, did not provide sufficient applanation pressure on the artery, leading to suboptimal measurement results.

In this study, we utilized a flexible piezoresistive yarn (FPY) pressure sensor, developed using a low-cost conductive thermoplastic polyurethane (TPU) filament as the base material. The sensor design was optimized based on the applanation tonometry principles, resulting in enhanced sensitivity and the ability to detect low-magnitude ABP waveforms in the radial artery. The sensor design facilitates precise and steady placement, ensuring that the continuous ABP waveform detected is both accurate and stable.

\section{Materials and Methods}
\subsection{Measuring Devices}
This study used an FPY pressure sensor based on conductive TPU filament containing conductive particles to measure ABP waveform. The base material for this sensor is a low-cost conductive filament (Recreus, Alicante, Spain) processed by a simple extrusion method using a 3D printer. This sensor detects pressure based on the deformation of the conductive particle structure caused by strain or pressure applied to its surface. This sensor has been previously evaluated for measuring various physiological signals in the human body, using two distinct measurement modes (strain or pressure) based on the characteristics of each signal. Each measurement result demonstrated high morphological similarity to reference standard and low baseline drift during continuous measurements \citep{ref-23}.

\begin{figure}[t!]
\centering
\includegraphics[width=9.5 cm]{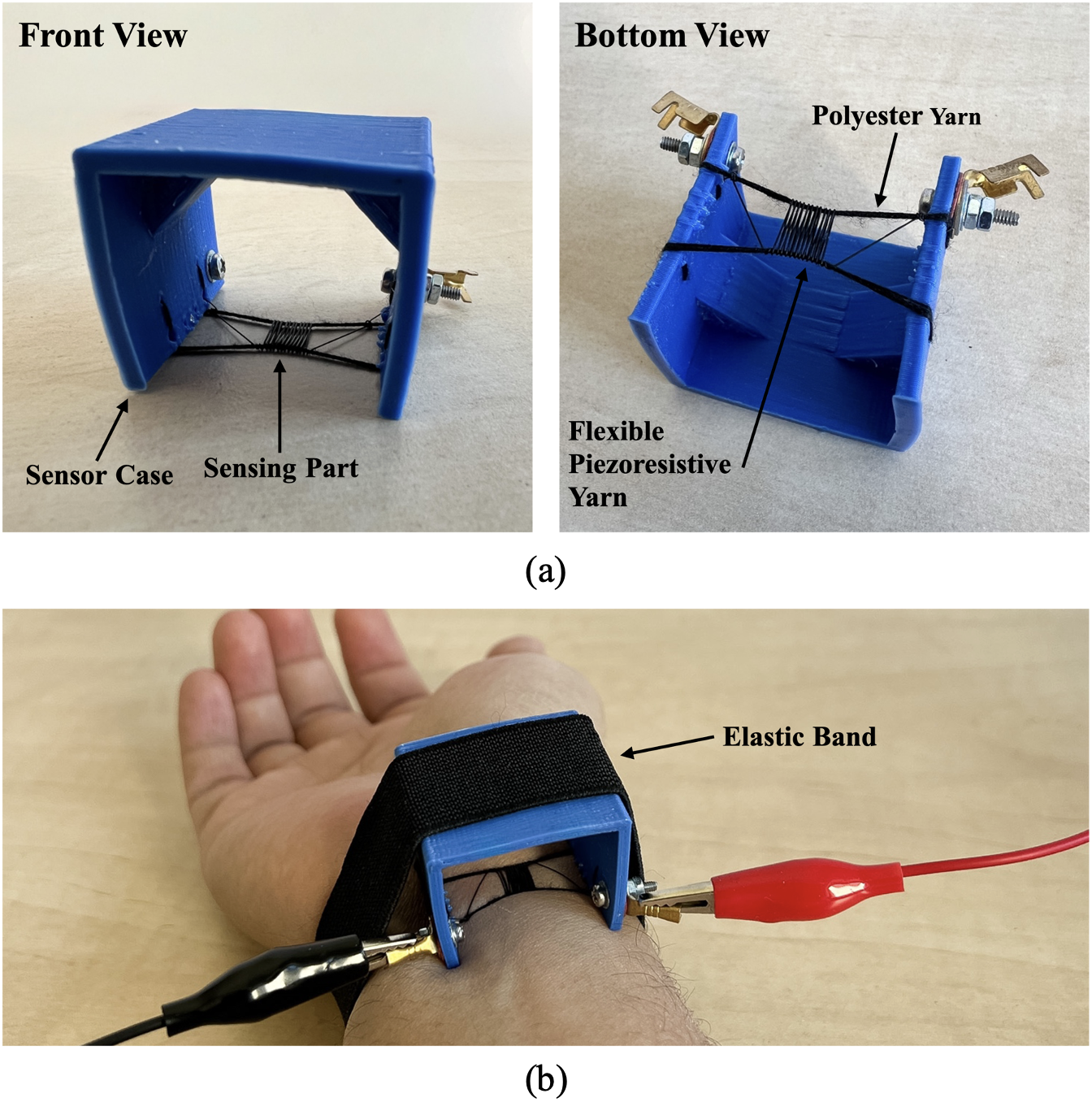}
\caption{FPY pressure sensor. (\textbf{a}) Front and bottom view. (\textbf{b}) Attached to the wrist using an elastic band.\label{fig1}}
\end{figure}

In this study, the FPY sensor operating in pressure mode is used, comprising two main components: the sensing part and the sensor case. The sensing part is assembled from FPY bonded to two parallel polyester yarns, forming a triangular pattern between them. The sensing part is sufficiently compact to cover the measurement point on the arterial site. The sensor case is designed as a hollow cube with open front, rear, and bottom sides. The sensing part is integrated into the lower center of the sensor case, as shown in Figure~\ref{fig1}a. During implementation, the sensor is attached to the wrist at the radial artery site using an elastic band, as shown in Figure~\ref{fig1}b.

The principle of pressure measurement in this sensor is based on the percolation concept in conductive particle structures on the FPY. Under initial conditions, conductive particles establish a conductive network with a baseline electrical resistance. When pressure is applied to the FPY surface, the conductive particles therein disperse, thereby diminishing the conductive network and increasing electrical resistance. Meanwhile, a decrease in electrical resistance involves the inverse process.

The designed sensor is ideal for arterial pressure measurement, applying the applanation tonometry method. The applanation process within the sensor integrates the feature of the sensing part and the sensor case. The flexibility of the sensing parts, primarily FPY and polyester yarn, enables them to conform to the curvature of the wrist. Securing the sensor case to the wrist with an elastic band enables the two polyester yarns to apply light pressure on the radial artery and flatten it. The area between the two polyester yarns, where the FPY bonding pattern forms, is the optimal location for measuring arterial pressure. During the measurement process, periodic arterial pressure continuously generates pressure on the sensing part in each cardiac cycle. This pressure causes deformation of the sensor and subsequently changes its electrical resistance.

The advantage of this sensor lies in its rubber FPY material, which is both flexible and comfortable to wear. The primary concern in implementing this sensor, as with other applanation tonometry-based sensors, is the precision of sensor placement on the radial artery. The tightness of the elastic band must also be considered to ensure that the sensing part provides adequate applanation pressure to the radial artery. Furthermore, adequate tightness ensures the sensor's position remains stable during measurement.

\subsection{Validation Devices}
Measurements of the ABP waveform in this study were performed using two scenarios. The first scenario involved using a hardware simulator to evaluate the sensor's response to a consistent waveform. The second scenario involved direct measurements at the radial artery site on the wrist.

\subsubsection{Measurement using Simulators}
\begin{figure}[t!]
\centering
\includegraphics[width=13.8 cm]{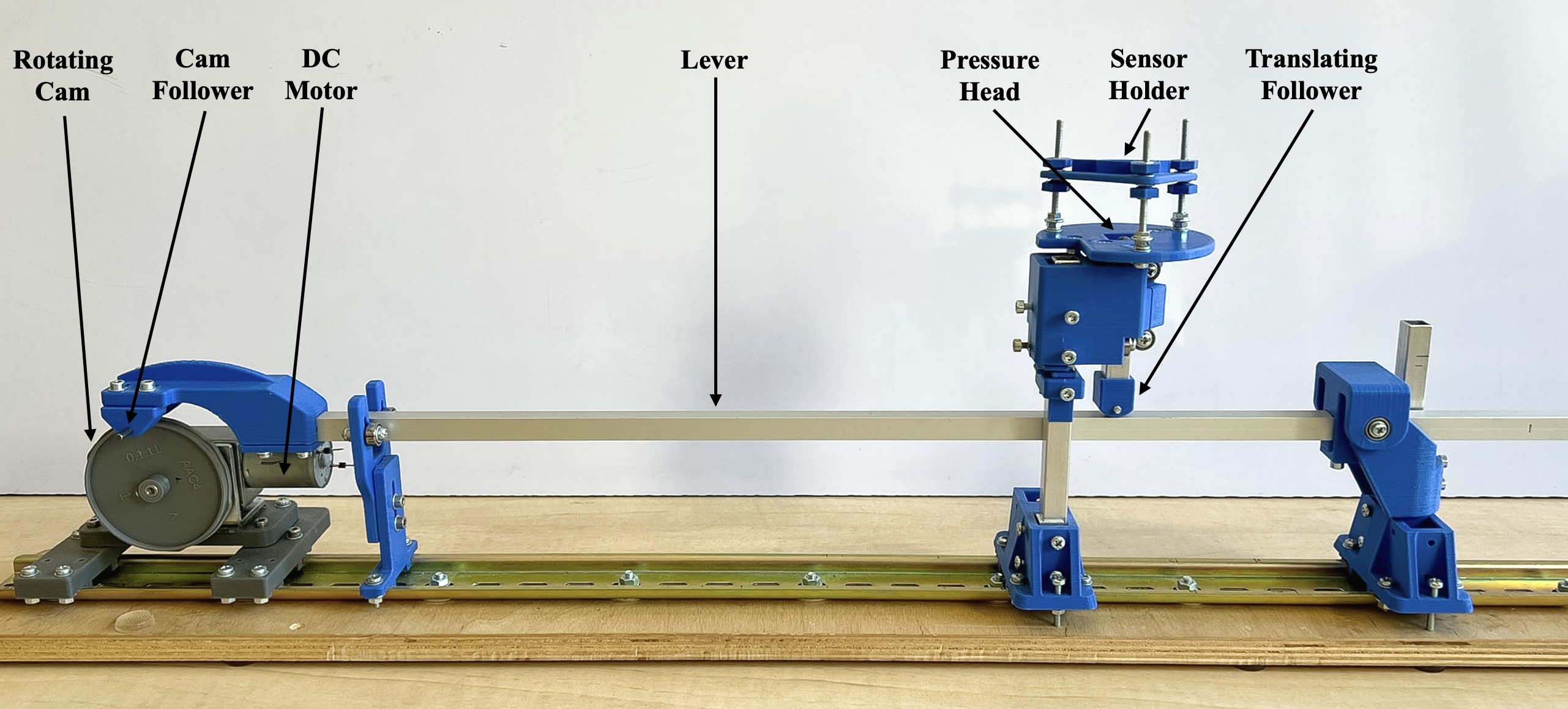}
\caption{The ABP waveform hardware simulator and its main components.\label{fig2}}
\end{figure}

The hardware simulator used in this study was developed by Répai et al. \citep{ref-24} and has been validated for its accuracy in generating ABP waveforms with an OptoForce OMD-20-SE-40N tactile force sensor. This simulator is designed as a tool for validating and evaluating continuous non-invasive blood pressure (CNIBP) sensors, primarily based on applanation tonometry. This device can simulate a continuous ABP waveform with high precision and offers several configurable parameters to optimize the testing of sensors with specific characteristics. The main components of this simulator include an interchangeable 3D-printed rotating cam, a cam follower, a DC motor, a lever, a translating follower, a pressure head and a sensor holder, as shown in Figure~\ref{fig2}. The 3D-printed cam is a crucial component for simulating the ABP waveform pattern. The 3D-printed cam features six complex contour sets on its surface, representing six continuous ABP waveform cycles.

The DC motor is used to rotate the 3D-printed cam attached to its shaft. To simulate a normal pulse rate (60 beats per minute), the motor rotates at 10 revolutions per minute (rpm). A cam follower, which is linked to a lever, is attached directly above the 3D-printed cam. This cam follower, made from stainless steel and having low friction against the 3D-printed cam, is designed to follow the contours of the rotating 3D-printed cam surface. This mechanism induces vertical movement along the lever in accordance with the contours of the 3D-printed cam surface. A translating follower is attached perpendicularly to the lever on the opposite side of the cam follower. The movement along the lever is translated into vertical movement only on the z-axis by the translating follower. At the top of the translating follower, there is an interchangeable pressure head that makes direct contact with the sensor surface. The sensor is positioned directly above the pressure head and secured in a sensor holder to provide sensor position stability during measurement. The simulator mechanism directly actuates the sensor via a planar linkage connected to a rotating 3D-printed cam. Moreover, the force magnitude applied to the sensor can be controlled by adjusting the distance between the 3D-printed cam and the translating follower.

The use of a simulator in this study aims to evaluate sensor performance, specifically focusing on repeatability, reproducibility, and accuracy in controlled measurement conditions.

\subsubsection{Measurement at the Arterial Site}
The OptoForce OMD-20-SE-40N tactile force sensor was used as a validation device for ABP waveform measurements at arterial sites. The sensor features a hemispherical dome surface composed of silicon rubber. Within the dome, a single light-emitting diode (LED) and four light-sensing elements are configured. During sensor operation, the LED emits light directed onto the dome's upper inner surface, where it is reflected, and a light-sensing element measures the reflected light intensity. The sensor detects pressure based on the deformation of its surface: pressure induces surface deformation, and the magnitude of this deformation affects the amount of reflected light intensity measured by each light-sensing element. Földi et al. \citep{ref-25} demonstrated the reliability of this sensor in measuring ABP waveforms at arterial sites. This sensor was chosen as a validation device because its working principle is based on applanation tonometry, as is our proposed sensor.

\subsection{Measurement Protocol}
\begin{figure}[t!]
\centering
\includegraphics[width=6.5 cm]{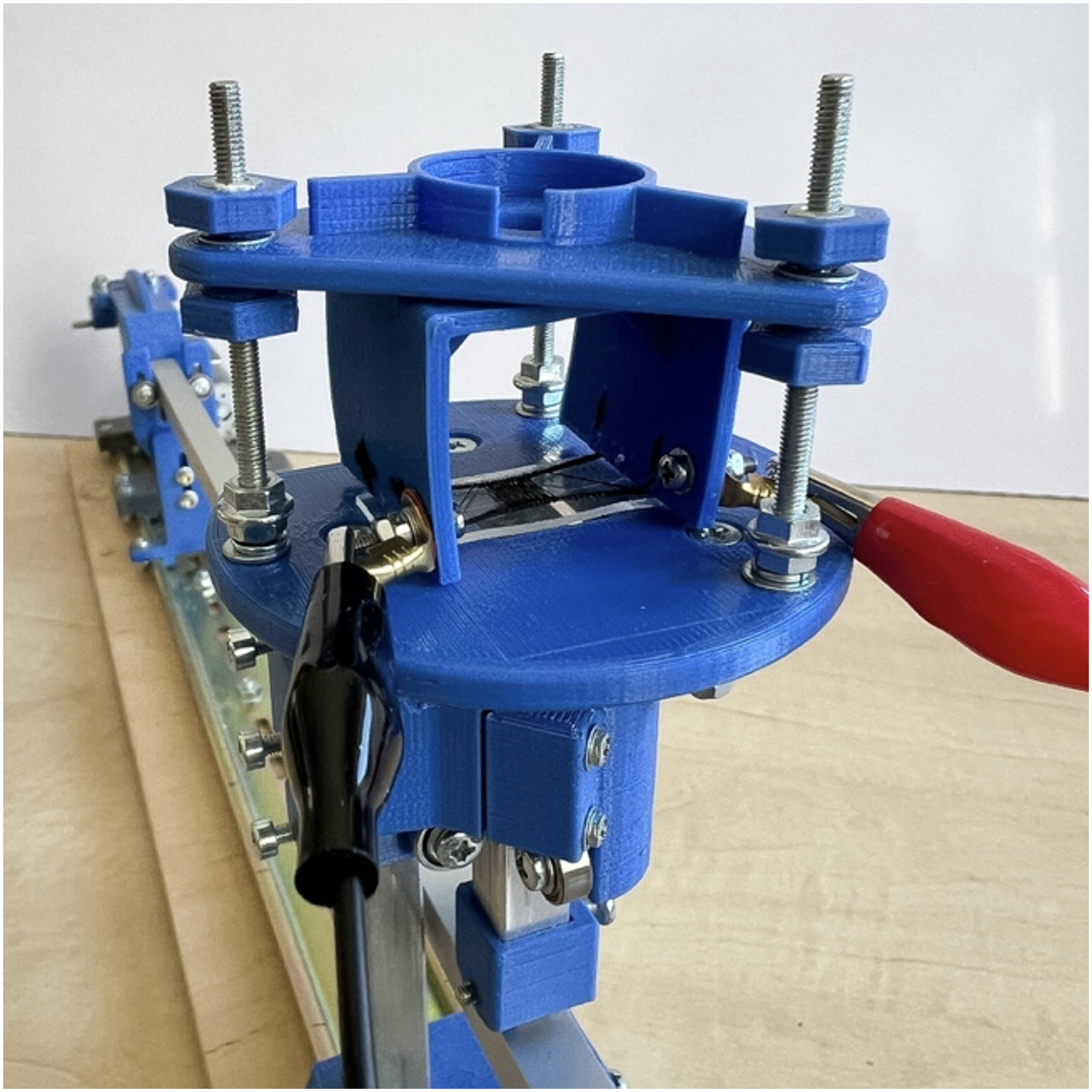}
\caption{The placement of the FPY pressure sensor on the sensor holder in the simulator.\label{fig3}}
\end{figure}

\begin{figure}[t!]
\centering
\includegraphics[width=10 cm]{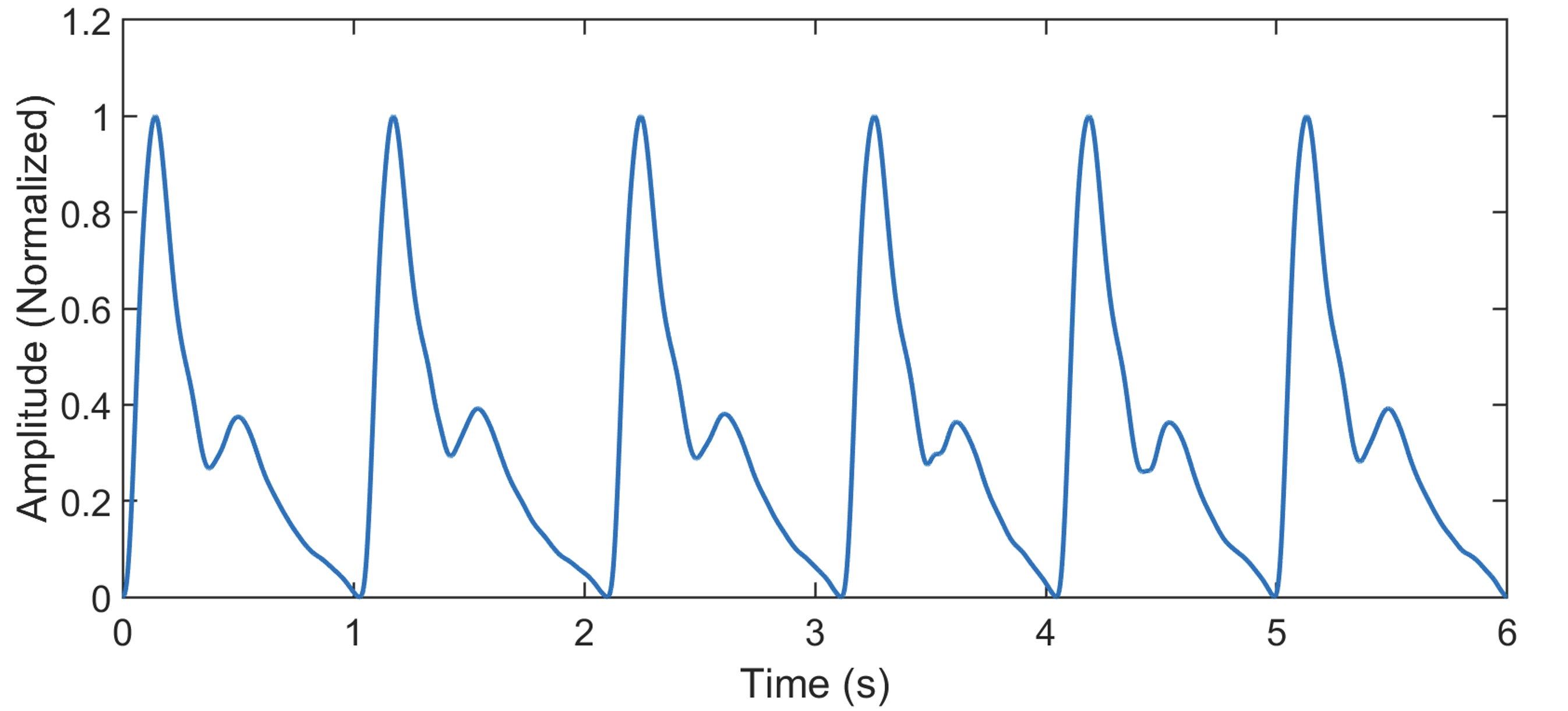}
\caption{Six consecutive ABP waveform cycles used as input to the simulator, corresponding to one full cam rotation.\label{fig4}}
\end{figure}

First, for the measurements using the simulator, the configurations and parameters used are based on the study by Répai et al. \citep{ref-24}, which aimed to generate realistic forces corresponding to arterial pressure. The measurements were performed at two different force magnitudes by adjusting the distance between the 3D-printed cam and the translating follower. A lower force of approximately 0.35 N is generated when the distance between the 3D-printed cam and the translating follower is set to 50 cm, whereas a higher force of approximately 0.75 N is generated when the distance is reduced to 40 cm. The placement of the FPY pressure sensor on the sensor holder in the simulator is shown in Figure~\ref{fig3}. A Polydimethylsiloxane (PDMS) layer was added between the FPY pressure sensor and the pressure head to mimic the function of the epidermal tissue at the arterial site on the wrist. A hemispherical pressure head with an 8 mm diameter was used to adjust the contact between the sensor surface and the pressure head. The simulated ABP waveform used in the measurement was collected from the PhysioNet database titled "Autonomic Aging: A dataset to quantify changes of cardiovascular autonomic function during healthy aging" \citep{ref-26} for the 30-34 year age group with Record ID 0004. Six consecutive initial signal cycles were selected and subsequently printed as a 3D cam. The ABP waveform corresponding to one full rotation of the 3D-printed cam is shown in Figure~\ref{fig4}. Each measurement session at each force level lasted 60 seconds and was repeated 10 times. Measurements at higher force are conducted after all lower force measurement sessions are complete.

Additional measurements were conducted to evaluate the sensor’s reproducibility. The configurations and parameters used in this measurement aligned with previous descriptions, with measurements recorded on three different days. Each measurement session began with reattaching the sensor to the simulator. A single measurement session at each force level was performed for 60 seconds daily.

A real ABP waveform was measured in a single healthy 36-year-old subject (the author). All procedures were non-invasive and performed in accordance with all relevant rules and regulations. Data collection and publication were conducted with informed consent.

Following a precise attachment of the sensor to the radial artery site and appropriate tightening of the elastic band, measurement can be performed with the subject in a seated position. The subject's wrist was placed on the table, and they were prohibited from making excessive movements during the measurement. Measurements were taken on the subject's dominant hand, which in this case is the right hand. Each measurement session lasted 30 seconds and was conducted 10 times for both the FPY pressure sensor and the OptoForce sensor. Measurements with the OptoForce sensor are conducted after all the FPY pressure sensor measurement sessions are complete.

All measurements using the FPY pressure sensor were conducted at the sampling frequency of 500 Hz. The resistance output from the FPY pressure sensor is converted into voltage using a voltage divider circuit with a 1 M{$\Omega$} pull-up resistor. Data recording from the FPY sensor measurements is performed using an oscilloscope (Keysight, California, USA).

\subsection{Data Analysis}
The signal processing for ABP waveform measurement using an FPY pressure sensor begins with three filtering stages: a notch filter, a low-pass filter, and a moving average filter. The notch filter is designed to remove 50 Hz noise caused by power-line interference. A low-pass filter with a cut-off frequency of 20 Hz is applied to eliminate high-frequency noise. Then, a moving average filter with a window size of 15 data points is used to remove motion artefacts. Initial testing indicates that this window size is optimal for removing motion artefacts while preserving the accuracy of waveform information in the signal.

For sensor validation, specifically for measurements using a simulator, approximately 42 seconds of the middle portion of the signal is selected from a 60-second measurement. The selected signal contains the ABP waveform from seven full cam rotations, starting with the cycle corresponding to the marked starting point of the cam. Whereas, for measurement at the arterial site, both using the FPY pressure sensor and the OptoForce sensor, approximately 20 seconds of the middle portion of the signal is selected from a 30-second measurement. This method is used to eliminate potential noise at the beginning and at the end of each measurement. The selected signal is then normalized to an amplitude range of 0 to 1, using the minimum and maximum amplitudes from each measurement procedures (lower force, higher force, or wrist measurement) as references.

The primary validation process is conducted using an average single-period signal to effectively reduce residual random noise, thereby providing a stable, representative waveform for subsequent analysis. Therefore, it is required to segment each single-period signal in the selected signal. Segmentation is performed by finding the local minimum as the onset point of each single-period signal. The following step is to compute the average of the single-period signal. Since the duration of each single-period signal varies, the signals are truncated to the shortest duration prior to averaging.

\subsection{Statistical Methods}
The repeatability of the FPY pressure sensor can be evaluated using measurements from a simulator, as it can generate consistent input waveforms for each full cam rotation. For this evaluation, the previously selected signals from the simulator measurements are used. During a single measurement session, the waveform detected over seven full cam rotations needs to be segmented into six ABP waveform cycles per rotation. Then, the average single-period signal for each rotation is computed. Correlation and root mean square error (RMSE) were used to evaluate the repeatability of this sensor, and their values were calculated for all averaged single-period signals between each rotation in the same measurement session. A high correlation and low RMSE value indicate that the FPY pressure sensor has good repeatability. The average correlation value for the selected signal in a single measurement session is calculated using Equation~\ref{eq1}.

\begin{equation}
    corr\_rpt_{avg} = \frac{corr\_rpt_{1-2} + corr\_rpt_{1-3} + \cdots + corr\_rpt_{6-7}}{21},\label{eq1}
\end{equation}
where {$corr\_rpt_{1-2}$} is the correlation value between the first and second rotations of the cam within the same measurement session, {$corr\_rpt_{1-3}$} is the correlation value between the first and third rotations, and so forth. The average RMSE value for a single measurement session is also calculated using the same method.

Reproducibility can be evaluated by comparing measurement results from the simulator on three different days using a similar approach. The difference lies in the calculation of the average single-period signal. Rather than using the average of each rotation, which represents repeatability, it uses the average of the single-period signal from a single measurement session (single day). Correlation and RMSE were calculated for all averaged single-period signals between each day. The average correlation values for the comparison of daily measurements are calculated using Equation~\ref{eq2}.

\begin{equation}
    corr\_rpd_{avg} = \frac{corr\_rpd_{1-2} + corr\_rpd_{1-3} + corr\_rpd_{2-3}}{3},\label{eq2}
\end{equation}
where {$corr\_rpd_{1-2}$} is the correlation value between the first and the second day measurement, {$corr\_rpd_{1-3}$} is the same for the first and the third day, and {$corr\_rpd_{2-3}$} is the same for the second and the third day. The average RMSE value is also calculated using the same method.

To evaluate the accuracy of ABP waveform measurement using a simulator, correlation and RMSE were calculated between the sensor measurements and the reference signal used to fabricate the 3D-printed cam. The measurement results used for the repeatability test were also used in this evaluation. As mentioned previously, a full rotation of the 3D-printed cam includes six ABP waveform cycles. The averaged single-period signal from each rotation in each measurement session is compared with the averaged single-period signal of the reference signal. The average correlation and RMSE value for each measurement session were calculated, along with their standard deviations, to assess variability.

Correlation and RMSE were also used to evaluate the accuracy of ABP waveform measurements at the arterial site by comparing FPY pressure sensor measurements with OptoForce sensor measurements. The selected signals, each roughly 20 seconds in duration from each measurement session in both sensors, were used in this validation, and the average single-period signal for each is calculated. The comparison method differed because the measurements from the two sensors were not acquired simultaneously. Our strategy was to compare the average single-period signal from each measurement session using the FPY pressure sensor with the average single-period signal from each measurement session using the OptoForce sensor. Then, as in the simulator measurement, the average correlation and RMSE value for each FPY pressure sensor measurement session were calculated, along with their standard deviations.

\begin{figure}[t!]
\centering
\includegraphics[width=11 cm]{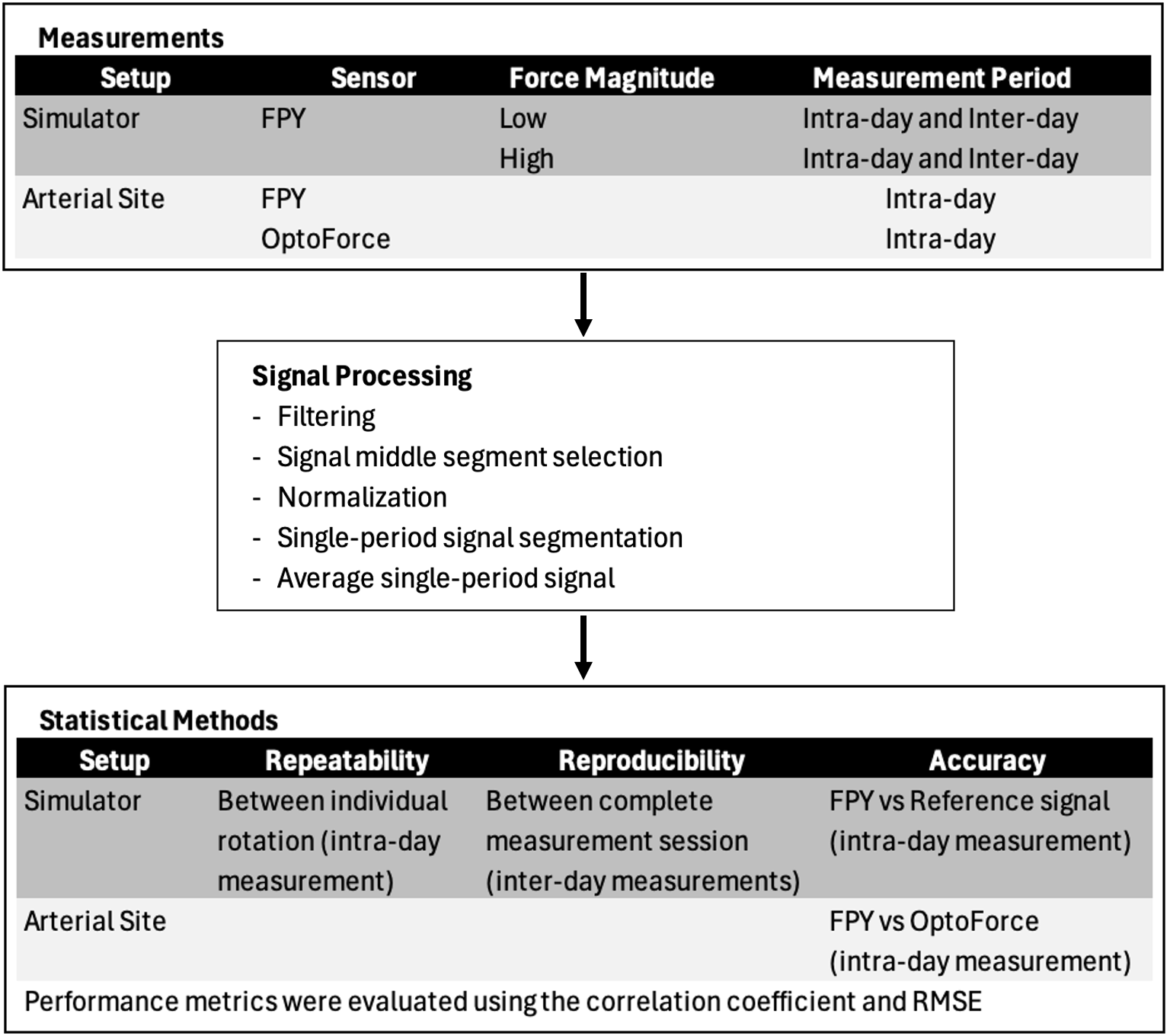}
\caption{A summary of all measurements, signal processing, and statistical methods in our study.\label{fig5}}
\end{figure}

A summary of all measurements, signal processing steps, and statistical methods is presented in Figure~\ref{fig5}.

\section{Results and Discussion}
\subsection{Measurement using Simulators}
\subsubsection{Repeatability Test}
\begin{figure}[t!]
\centering
\includegraphics[width=10 cm]{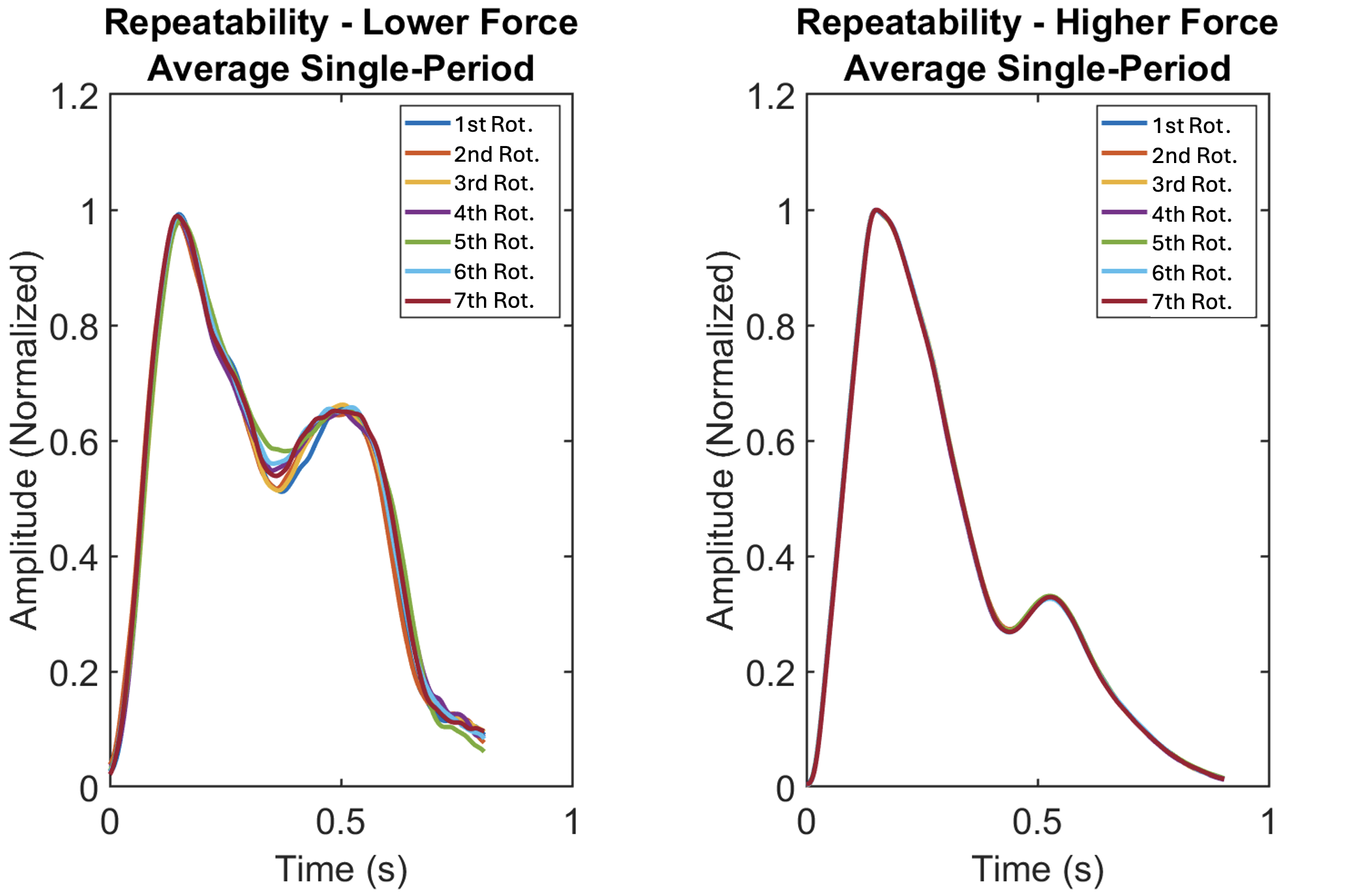}
\caption{The averaged single-period signal from each rotation during a single measurement session, with the best repeatability for each measurement at lower force (correlation = 0.9965) and higher force (correlation= 0.99997).\label{fig6}}
\end{figure}

The results of the repeatability test using the average single-period signal for both lower and higher forces in each measurement session are presented in Table~\ref{tab1}. The correlation values in these tables were calculated using Equation~\ref{eq1}, and the RMSE values were calculated using the same approach by averaging the RMSE values from the comparison of each rotation during the same measurement session. For measurement using lower force, the average correlation was 0.9896 ± 0.0096 with an average RMSE of 0.0409 ± 0.0165. Meanwhile, for measurement using higher force, the average correlation was 0.9999 ± 0.0002 with an average RMSE of 0.0049 ± 0.0022. The average single-period signal for each rotation from a single measurement session with the best correlation at each of the two force levels is shown in Figure~\ref{fig6}. The repeatability tests revealed that lower force measurements had slightly lower correlation values and slightly higher RMSE values compared to higher force measurements. This variation occurs because measurement noise has a more significant impact on low-magnitude signal measurement. In this test, measurement noise may arise from fabrication defects in the 3D-printed cam or vibrations generated by the DC motor during rotation. A detailed discussion of the morphological differences between the lower and higher force measurements is provided in the accuracy test section.

\begin{table}[t!]
\centering
\caption{Repeatability test results using the averaged single-period signal at lower and higher forces on the hardware simulator.}
\label{tab1}
\begin{tabular}{lllll}
\toprule
& \multicolumn{2}{l}{\textbf{Lower Force}} & \multicolumn{2}{l}{\textbf{Higher Force}} \\
\cmidrule(l){2-3} \cmidrule(l){4-5}
& \textbf{Correlation} & \textbf{RMSE} & \textbf{Correlation} & \textbf{RMSE} \\
\midrule
Meas 01 & 0.9822 $\pm$ 0.0151 & 0.0498 $\pm$ 0.0216 & 0.99997 $\pm$ 0.00002 & 0.0030 $\pm$ 0.0007 \\
Meas 02 & 0.9892 $\pm$ 0.0089 & 0.0428 $\pm$ 0.0140 & 0.99988 $\pm$ 0.00008 & 0.0050 $\pm$ 0.0016 \\
Meas 03 & 0.9867 $\pm$ 0.0142 & 0.0440 $\pm$ 0.0202 & 0.99991 $\pm$ 0.00008 & 0.0046 $\pm$ 0.0016 \\
Meas 04 & 0.9879 $\pm$ 0.0093 & 0.0443 $\pm$ 0.0146 & 0.99995 $\pm$ 0.00003 & 0.0038 $\pm$ 0.0012 \\
Meas 05 & 0.9923 $\pm$ 0.0053 & 0.0408 $\pm$ 0.0162 & 0.99977 $\pm$ 0.00023 & 0.0058 $\pm$ 0.0031 \\
Meas 06 & 0.9909 $\pm$ 0.0047 & 0.0396 $\pm$ 0.0111 & 0.99986 $\pm$ 0.00011 & 0.0048 $\pm$ 0.0020 \\
Meas 07 & 0.9965 $\pm$ 0.0024 & 0.0243 $\pm$ 0.0071 & 0.99984 $\pm$ 0.00014 & 0.0055 $\pm$ 0.0022 \\
Meas 08 & 0.9953 $\pm$ 0.0028 & 0.0279 $\pm$ 0.0078 & 0.99987 $\pm$ 0.00010 & 0.0048 $\pm$ 0.0019 \\
Meas 09 & 0.9857 $\pm$ 0.0096 & 0.0479 $\pm$ 0.0147 & 0.99978 $\pm$ 0.00021 & 0.0060 $\pm$ 0.0027 \\
Meas 10 & 0.9893 $\pm$ 0.0052 & 0.0472 $\pm$ 0.0137 & 0.99981 $\pm$ 0.00022 & 0.0056 $\pm$ 0.0028 \\
\bottomrule
\end{tabular}
\end{table}

\begin{figure}[t!]
\centering
\includegraphics[width=11.5 cm]{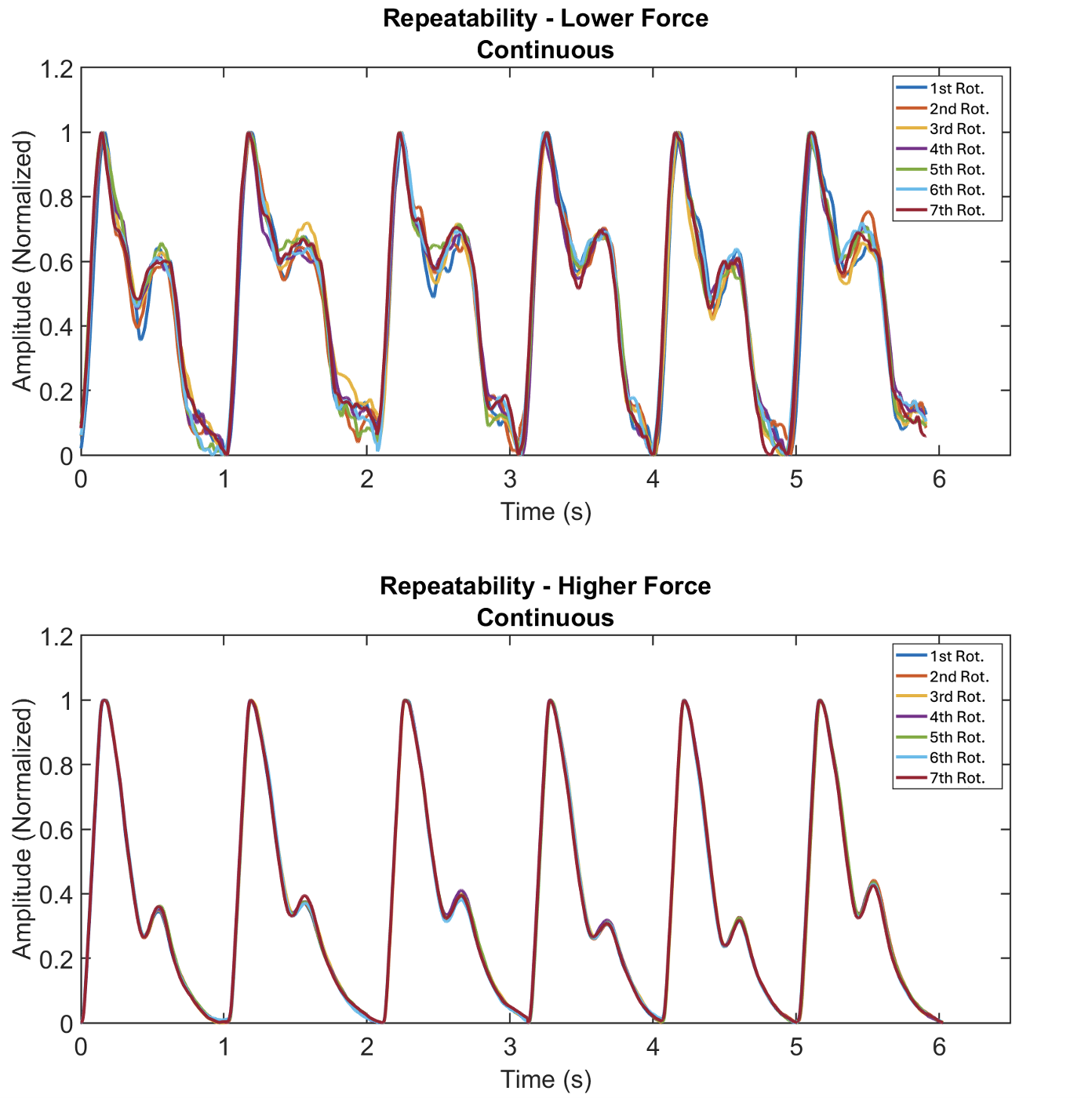}
\caption{The continuous signal from each rotation during a single measurement session, with the best repeatability for each measurement at lower force (correlation = 0.9877) and higher force (correlation= 0.9996).\label{fig7}}
\end{figure}

In addition to average single-period signal comparison, repeatability was also analyzed by comparing continuous signals from each rotation containing six continuous ABP waveform cycles. For measurement using lower force, the average correlation was 0.9794 ± 0.0098 with an average RMSE of 0.0615 ± 0.0148. Meanwhile, for measurement using higher force, the average correlation was 0.9988 ± 0.0016 with an average RMSE of 0.0132 ± 0.0069. The continuous signal for each rotation from a single measurement session with the best correlation at each of the two force levels is shown in Figure~\ref{fig7}. Repeatability evaluations using a continuous signal demonstrated slightly decreased correlation values and increased RMSE values. The averaging process in the averaged single-period signal method effectively reduces measurement noise, resulting in better repeatability evaluations. Furthermore, continuous signal evaluations are highly sensitive to minor sensor response delays, even a slight delay in detecting one cycle shifts the onset points of subsequent cycles, further degrading the correlation and RMSE.

In this repeatability test, the correlation value represents the sensor’s consistency in detecting the input signal’s morphology, while the RMSE value indicates the variance between repeated measurements. Overall, the FPY pressure sensor demonstrates high repeatability across all experimental parameters, including varying force levels and signal evaluation methods. Notably, several measurement sessions showed correlation values approaching one, indicating that the output signal morphology remains identical during each rotation. These results are further supported by the low RMSE values, demonstrating that the absolute differences between repeated measurements are very small.

\subsubsection{Reproducibility Test}
A reproducibility test was conducted by performing a single measurement session for each force level daily for three consecutive days. Repeated measurements on the same day were excluded, as previous tests validated that the sensor has high repeatability in intra-day measurements. Therefore, the reproducibility test focused on comparing measurement results across varying conditions, represented by differences in measurement day, with the procedure of reattaching the sensor in the simulator.

\begin{figure}[t!]
\centering
\includegraphics[width=10 cm]{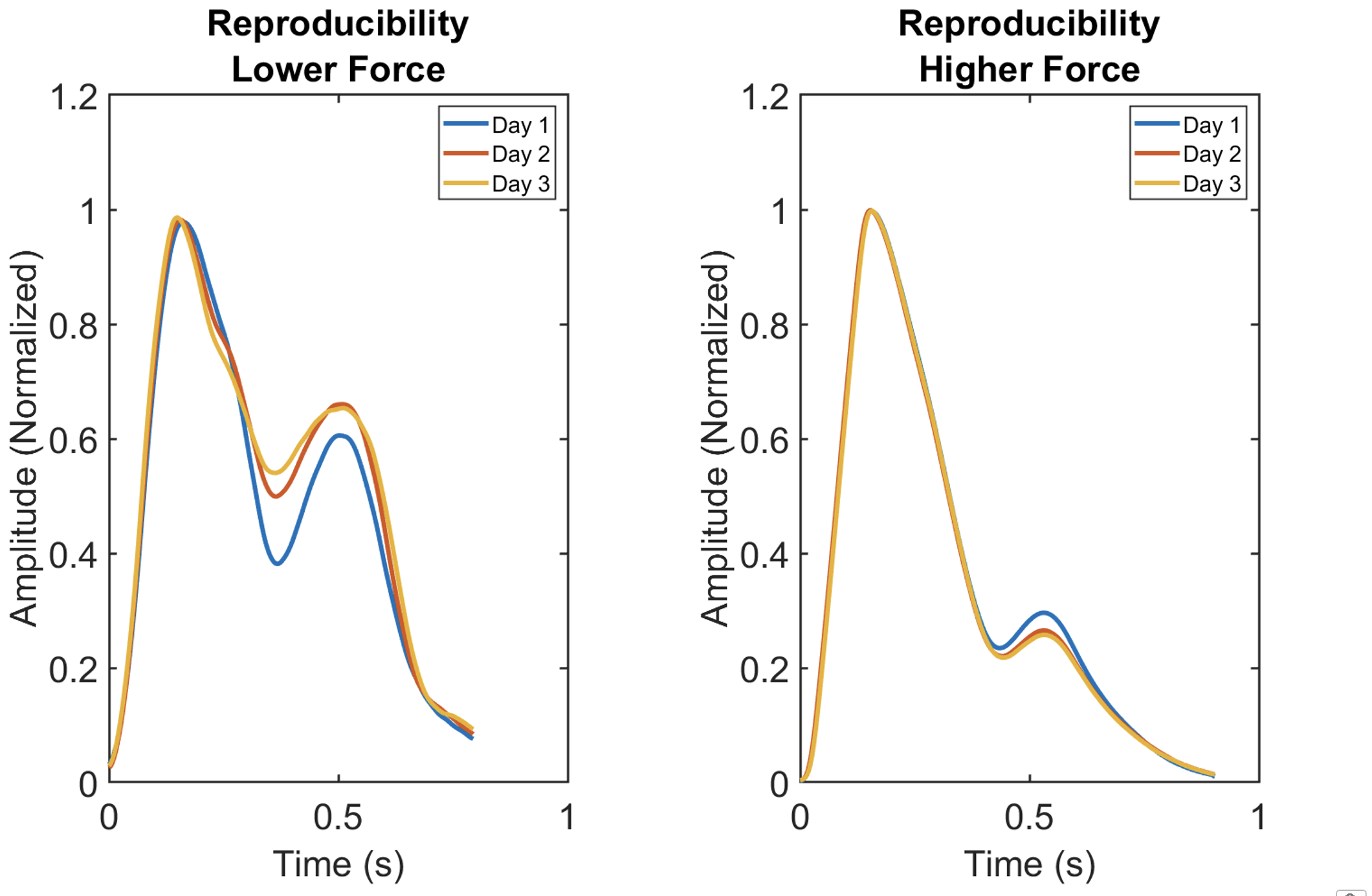}
\caption{The average single-period signal from a daily measurement session for lower force (correlation = 0.9879) and higher force (correlation = 0.9994).\label{fig8}}
\end{figure}

The average correlation values for the reproducibility test were calculated using Equation~\ref{eq2}, and the RMSE values were calculated using the same approach. Evaluations of the averaged single-period signal at lower forces resulted in an average correlation of 0.9879 ± 0.0095 with an average RMSE of 0.0498 ± 0.0243. Meanwhile, for measurements using higher force, the average correlation was 0.9994 ± 0.0002 with an average RMSE of 0.0125 ± 0.0037. The average single-period signal illustrating the correlation between inter-day measurements at both force levels is shown in Figure~\ref{fig8}.

In the context of continuous signal comparison, using all selected signals from a daily measurement session, the average correlation for measurements using lower force was 0.9617 ± 0.0104 with an average RMSE of 0.0871 ± 0.0133. Meanwhile, for measurements using higher force, the average correlation was 0.9762 ± 0.0137 with an average RMSE of 0.0650 ± 0.0187.

The correlation and RMSE values from the reproducibility tests show trends similar to those from the repeatability tests. Tests conducted at higher forces and evaluations using the average single-period signal resulted in higher correlation values and lower RMSE values. The overall results from the reproducibility tests demonstrate that the FPY pressure sensor operates reliably under varying test conditions, including instances where the sensor is reattached on the simulator. The correlation and RMSE values do not differ significantly from those obtained in tests conducted under static conditions (repeatability tests).

\subsubsection{Accuracy Test}
\begin{figure}[t!]
\centering
\includegraphics[width=13.8 cm]{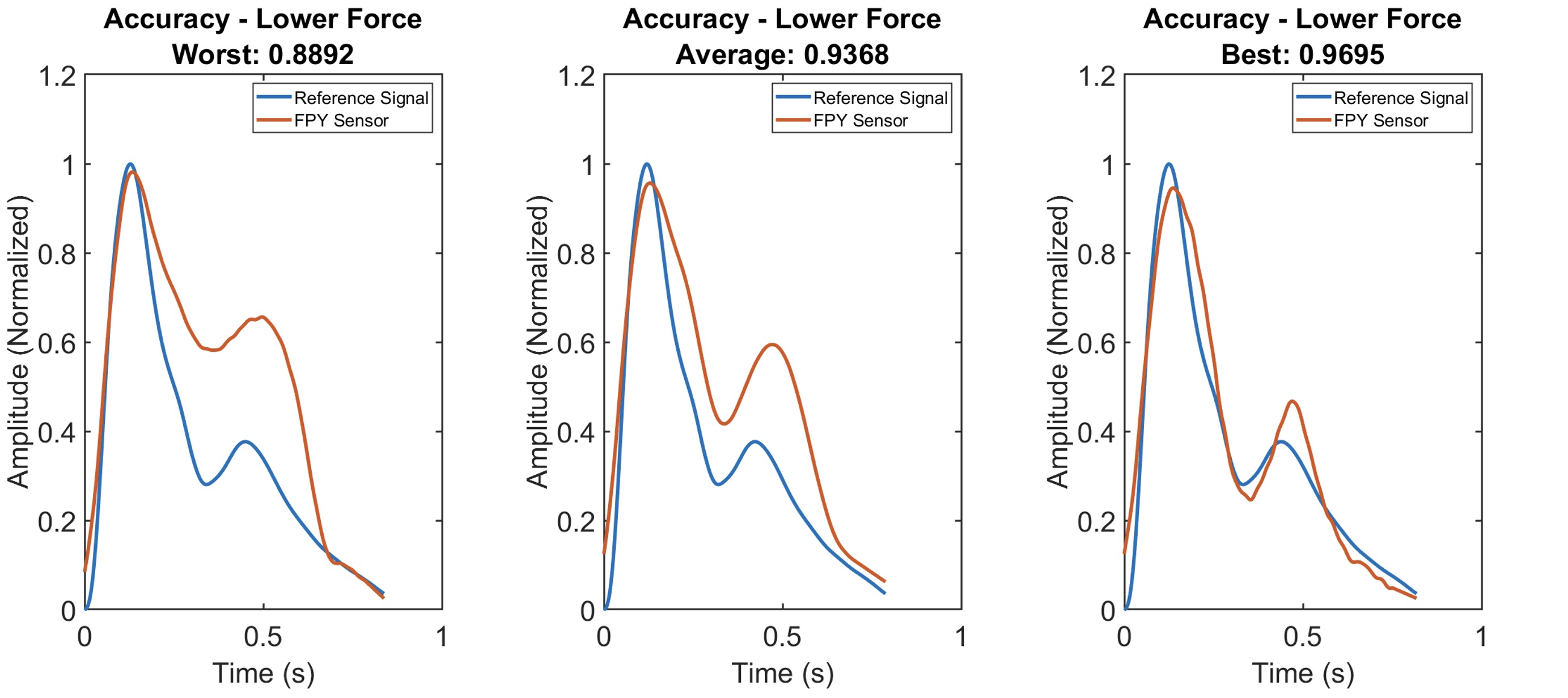}
\caption{The average single-period signal from specific rotation with the worst and best correlations and the grand-average single-period signal, compared to the reference signal at lower force.\label{fig9}}
\end{figure}

\begin{figure}[t!]
\centering
\includegraphics[width=13.8 cm]{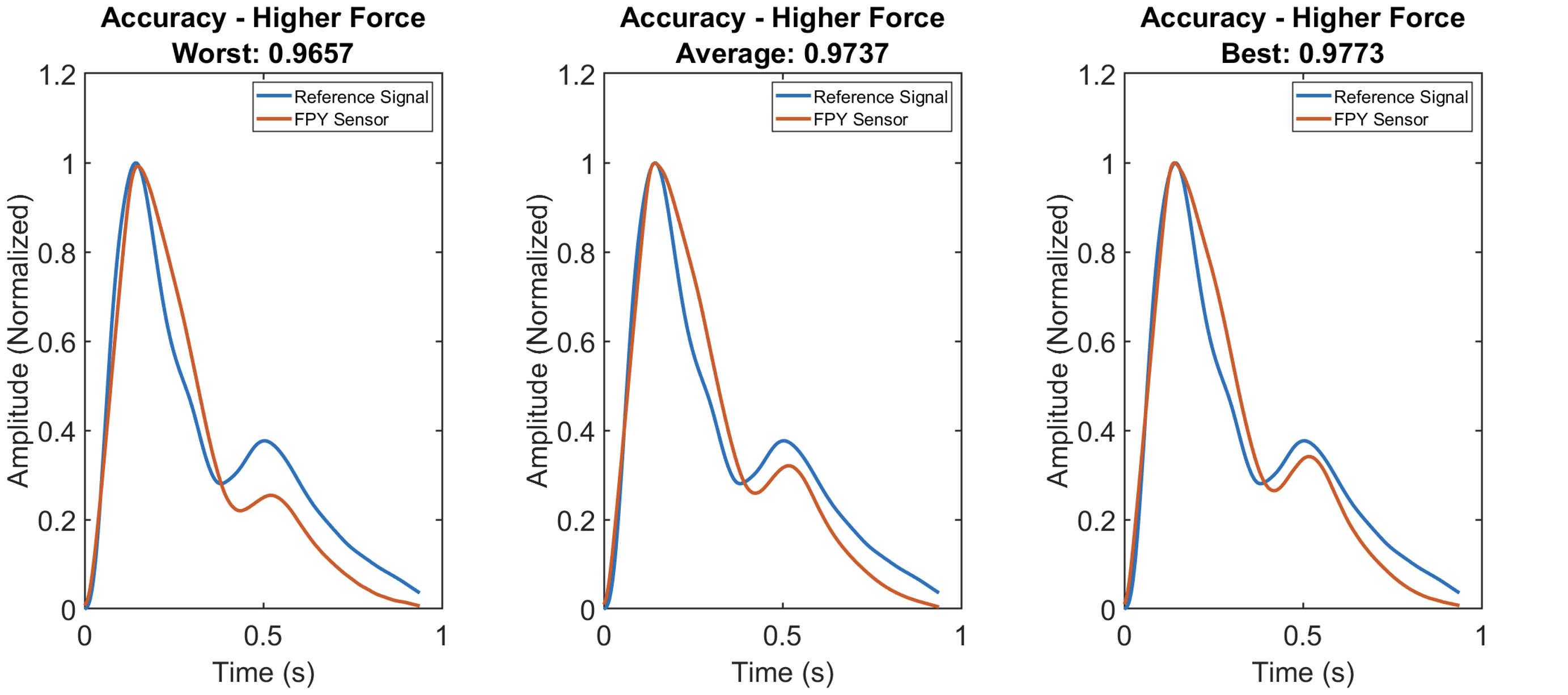}
\caption{The average single-period signal from specific rotation with the worst and best correlations and the grand-average single-period signal, compared to the reference signal at higher force.\label{fig10}}
\end{figure}

An accuracy test was conducted by comparing the measurement results from the FPY pressure sensor at each rotation with the reference signal. The results of the accuracy tests using the average single-period signal for lower and higher force in each measurement session are shown in Table~\ref{tab2}. The correlation and RMSE values for each measurement session in the table represent the averages obtained by comparing each rotation against the reference signal. In measurements using lower force, the average correlation was 0.9308 ± 0.0228 with an average RMSE of 0.1507 ± 0.0383. Meanwhile, in measurements using higher force, the average correlation increased to 0.9731 ± 0.0027, while the average RMSE decreased to 0.0769 ± 0.0044. The average single-period signal from specific rotations representing the worst and best correlation to the reference, along with the grand-average single-period signal derived from all recorded rotations compared to the reference, for the lower and higher force measurements, are shown in Figure~\ref{fig9} and Figure~\ref{fig10}, respectively.

\begin{table}[t!]
\centering
\caption{Accuracy test results using the averaged single-period signal at lower and higher forces on the hardware simulator.}
\label{tab2}
\begin{tabular}{lllll}
\toprule
& \multicolumn{2}{l}{\textbf{Lower Force}} & \multicolumn{2}{l}{\textbf{Higher Force}} \\
\cmidrule(l){2-3} \cmidrule(l){4-5}
& \textbf{Correlation} & \textbf{RMSE} & \textbf{Correlation} & \textbf{RMSE} \\
\midrule
Meas 01 & 0.9410 $\pm$ 0.0141 & 0.1423 $\pm$ 0.0148 & 0.9698 $\pm$ 0.0005 & 0.0813 $\pm$ 0.0007 \\
Meas 02 & 0.9442 $\pm$ 0.0101 & 0.1442 $\pm$ 0.0140 & 0.9730 $\pm$ 0.0011 & 0.0767 $\pm$ 0.0017 \\
Meas 03 & 0.9405 $\pm$ 0.0136 & 0.1433 $\pm$ 0.0143 & 0.9735 $\pm$ 0.0010 & 0.0758 $\pm$ 0.0013 \\
Meas 04 & 0.9289 $\pm$ 0.0170 & 0.1518 $\pm$ 0.0122 & 0.9731 $\pm$ 0.0008 & 0.0760 $\pm$ 0.0011 \\
Meas 05 & 0.9153 $\pm$ 0.0194 & 0.1859 $\pm$ 0.0229 & 0.9757 $\pm$ 0.0017 & 0.0719 $\pm$ 0.0021 \\
Meas 06 & 0.9243 $\pm$ 0.0163 & 0.1656 $\pm$ 0.0160 & 0.9757 $\pm$ 0.0010 & 0.0718 $\pm$ 0.0013 \\
Meas 07 & 0.9067 $\pm$ 0.0085 & 0.1878 $\pm$ 0.0083 & 0.9752 $\pm$ 0.0013 & 0.0737 $\pm$ 0.0014 \\
Meas 08 & 0.8963 $\pm$ 0.0044 & 0.2027 $\pm$ 0.0082 & 0.9751 $\pm$ 0.0008 & 0.0756 $\pm$ 0.0010 \\
Meas 09 & 0.9579 $\pm$ 0.0096 & 0.0852 $\pm$ 0.0098 & 0.9709 $\pm$ 0.0016 & 0.0811 $\pm$ 0.0017 \\
Meas 10 & 0.9531 $\pm$ 0.0100 & 0.0981 $\pm$ 0.0123 & 0.9687 $\pm$ 0.0015 & 0.0854 $\pm$ 0.0012 \\
\bottomrule
\end{tabular}
\end{table}

For the accuracy evaluations using continuous signals from each rotation, the average correlation for measurement using lower force was 0.8934 ± 0.0661, with an average RMSE of 0.1698 ± 0.0414. Meanwhile, for measurement using higher force, the average correlation increased to 0.9679 ± 0.0034, while the average RMSE decreased to 0.0818 ± 0.0044.

The correlation and RMSE values in the accuracy test also show a trend consistent with the repeatability and reproducibility tests. The correlation values at lower force, 0.9308 for the average single-period signal and 0.8934 for the continuous signal evaluations, are acceptable. Higher correlations were obtained in evaluations using higher force: 0.9731 for the average single-period signal and 0.9679 for the continuous signal evaluations.

A relatively high RMSE value (RMSE > 0.15) was obtained from measurements using a lower force, as depicted in the signal morphology shown in Figure~\ref{fig9}. This result is affected by the viscoelastic characteristics of the sensor’s base material, which behaves both as a thick liquid (viscous) and a solid spring (elastic). At higher forces, the sensor undergoes significant deformation, producing a large internal restoring force that dominates its viscous resistance. This condition causes the sensor’s resistance to return to its initial values faster once the pressure is released. On the other hand, at lower forces, the sensor undergoes less significant deformation, producing a weaker restoring force, making its viscous characteristics more dominant. This condition leads to delayed recovery of the sensor’s resistance to its initial value.

Based on the average single-period signal from the average correlation of measurements at lower force in Figure~\ref{fig9}, the amplitude of the dicrotic peak (second peak) in the FPY pressure sensor measurements is higher than that of the reference signal. In accordance with the previously described viscoelastic concept, this result is due to the delayed recovery in sensor resistance following the systolic peak (first peak). While the sensor resistance has not completely returned to its initial value, a subsequent pressure occurs, causing the amplitude of the dicrotic peak to be higher than its true value. The only measurement result at a lower force with a different amplitude of the dicrotic peak occurred during the last measurement session, which is one of the rotations in that session that has the best correlation with the reference signal. This concern may arise from a decrease in the sensor’s internal viscosity after prolonged use during the measurement period. As a result, this factor accelerated the recovery of the sensor’s resistance to its initial value.

Given the high correlation from the accuracy test, which indicates that the sensor can detect the main dynamic features of the ABP waveform, alongside the high repeatability, which indicates the consistency of the sensor, this allows for the morphological distortion caused by viscoelastic characteristics to be compensated and further processed by applying an inverse transfer function. Applying this inverse transfer function can reconstruct the amplitude of the dicrotic peak to align with the reference signal.

\subsection{Measurement at the Arterial Site}
Accuracy tests for ABP waveform measurements at the arterial site were performed by comparing the results of each measurement session from the FPY pressure sensor with those from the OptoForce sensor, which was used as the reference. It is important to understand that the OptoForce sensor is not the gold standard for ABP waveform measurement. However, this sensor demonstrated reliability and shows high correlation in ABP waveform measurement relative to measurements taken using a Millar tonometer \citep{ref-25}.

Measurements of the ABP waveform using the FPY pressure sensor and the OptoForce sensor are not performed simultaneously. Simultaneous measurements on the same wrist are unfeasible due to the narrow optimal measurement point at the arterial site. Furthermore, the applanation pressure exerted by a proximally positioned sensor directly distorts the pressure wave transmitted to the adjacent distal sensor. In addition, simultaneous measurements on different wrists do not ensure identical ABP waveforms due to variations in vascular anatomy at the two arterial sites. To evaluate the sensor’s accuracy in detecting the morphological characteristics of the ABP waveform, the sequential measurement method is considered appropriate. To minimize inter-time variability in the ABP waveform measurement, measurements using both sensors were conducted while the subject was resting to maintain stable blood pressure.

\begin{figure}[t!]
\centering
\includegraphics[width=10 cm]{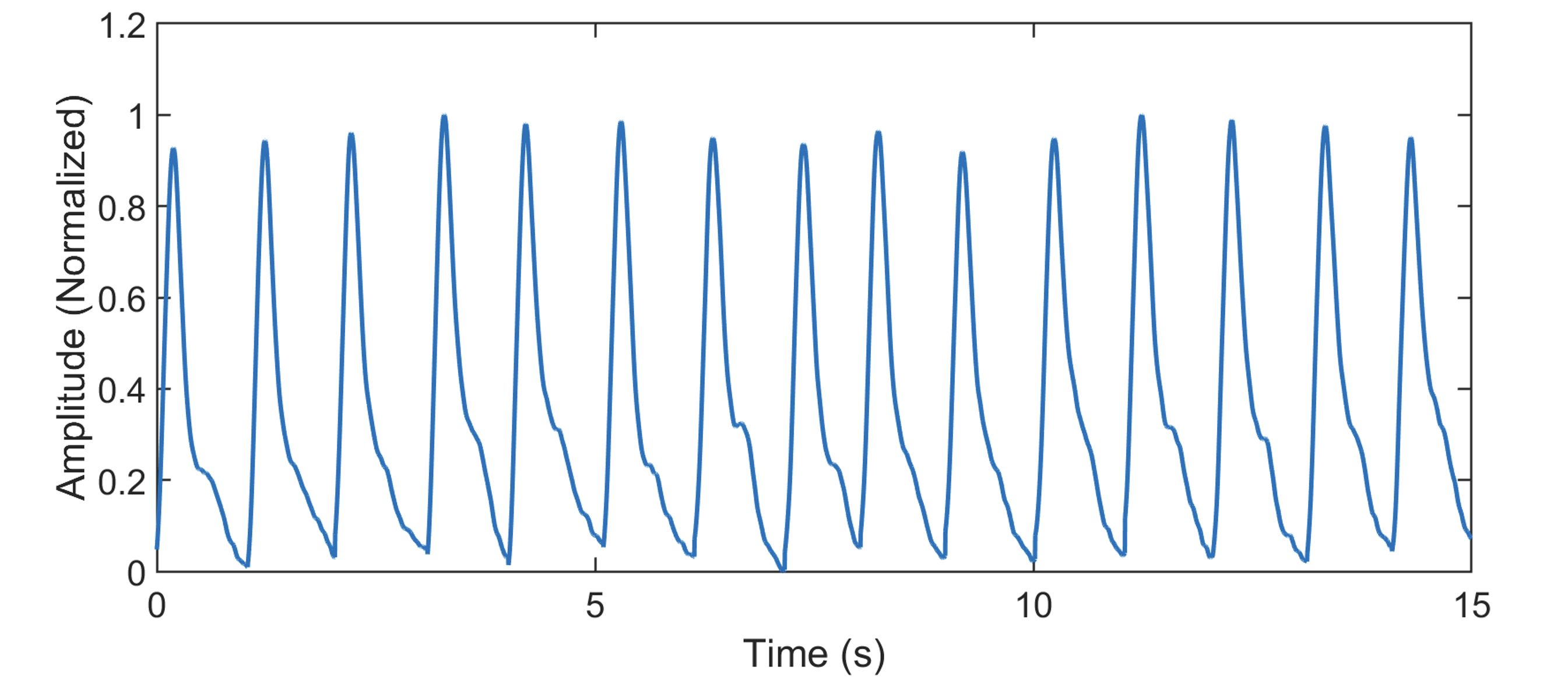}
\caption{A 15-second continuous ABP waveform measured by the FPY pressure sensor.\label{fig11}}
\end{figure}

\begin{figure}[t!]
\centering
\includegraphics[width=13.8 cm]{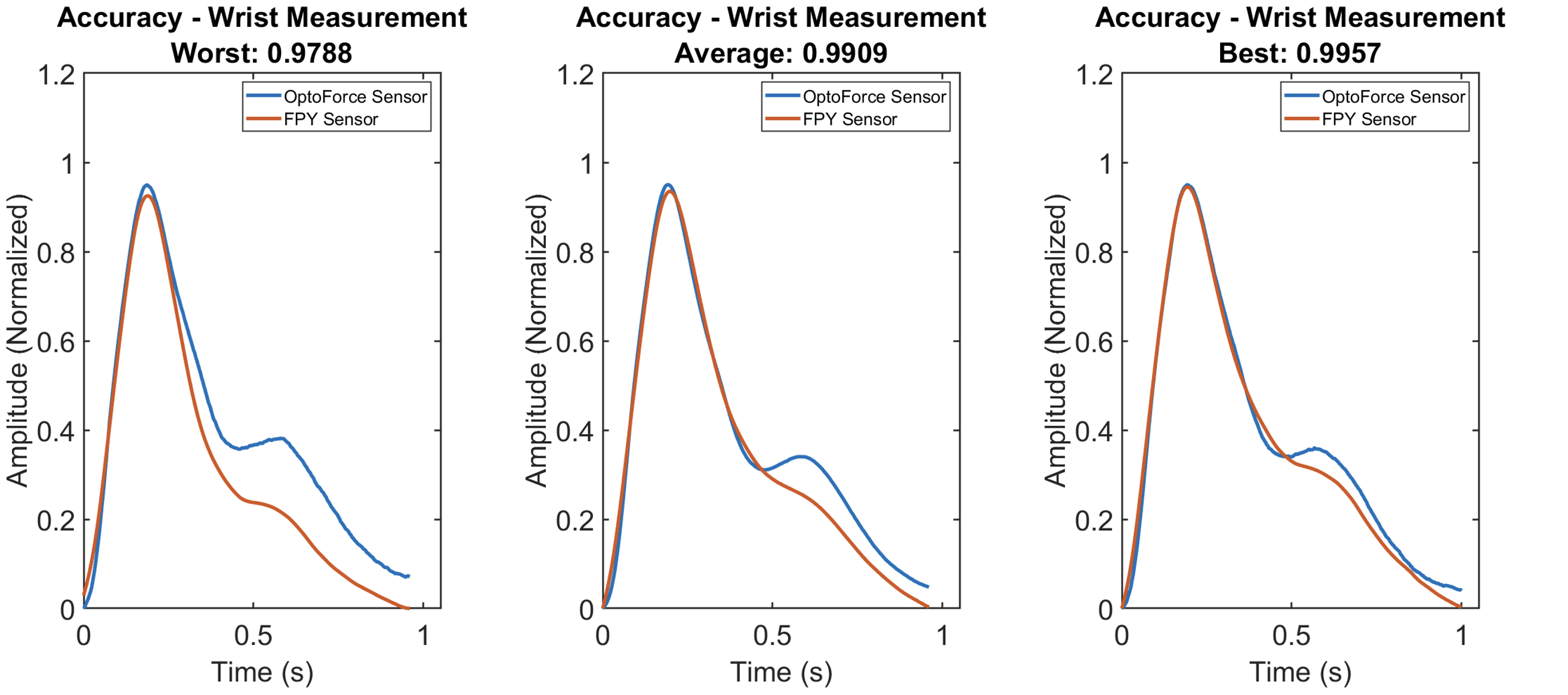}
\caption{The average single-period signal from paired measurement sessions of the FPY pressure sensor and the OptoForce sensor with the worst and best correlations and the average correlation across all measurement sessions.\label{fig12}}
\end{figure}

The accuracy test results using the average single-period signal for measurement at the arterial site are shown in Table~\ref{tab3}. The correlation and RMSE values in the table for each FPY pressure sensor measurement session represent the average correlation and RMSE from comparisons with each OptoForce sensor measurement session. The average correlation between the measurement results of these two sensors is 0.9880 ± 0.0035, with an average RMSE of 0.0665 ± 0.0160 for the FPY pressure sensor measurement relative to the OptoForce sensor measurement. The ABP waveform from a single measurement session using the FPY pressure sensor is shown in Figure~\ref{fig11}. The average single-period signal from paired measurement sessions of the FPY pressure sensor and the OptoForce sensor, representing the worst and best correlation, along with the average correlation over all measurement sessions, is shown in Figure~\ref{fig12}. An evaluation using continuous signals was not performed because the measurements were not taken simultaneously.

\begin{table}[t!]
\centering
\caption{Accuracy test results using the average single-period signal for measurements at the arterial site, comparing the FPY pressure sensor with the OptoForce sensor.}
\label{tab3}
\begin{tabular}{lll}
\toprule
& \textbf{Correlation} & \textbf{RMSE} \\
\midrule
Meas 01 & 0.9868 $\pm$ 0.0029 & 0.0581 $\pm$ 0.0075 \\
Meas 02 & 0.9885 $\pm$ 0.0022 & 0.0581 $\pm$ 0.0100 \\
Meas 03 & 0.9847 $\pm$ 0.0028 & 0.0759 $\pm$ 0.0148 \\
Meas 04 & 0.9854 $\pm$ 0.0021 & 0.0700 $\pm$ 0.0114 \\
Meas 05 & 0.9918 $\pm$ 0.0023 & 0.0506 $\pm$ 0.0092 \\
Meas 06 & 0.9898 $\pm$ 0.0027 & 0.0662 $\pm$ 0.0152 \\
Meas 07 & 0.9897 $\pm$ 0.0021 & 0.0613 $\pm$ 0.0111 \\
Meas 08 & 0.9881 $\pm$ 0.0035 & 0.0736 $\pm$ 0.0168 \\
Meas 09 & 0.9904 $\pm$ 0.0018 & 0.0648 $\pm$ 0.0137 \\
Meas 10 & 0.9846 $\pm$ 0.0040 & 0.0871 $\pm$ 0.0170 \\
\bottomrule
\end{tabular}
\end{table}

The measurement results from the FPY pressure sensor and the OptoForce sensor show a high correlation, with an acceptable RMSE value indicating a small difference between the actual values of the two measurements. Considering that the two measurements were not taken simultaneously, the waveform variations due to the time difference in signal acquisition contributed to the calculated RMSE value. In terms of signal morphology, the ABP waveform detected by the FPY pressure sensor shows a high similarity to the ABP waveform detected by the OptoForce sensor, especially at the systolic phase. During the diastolic phase, however, a minor morphological deviation is observed, as the FPY pressure sensor detects an attenuated dicrotic peak. This phenomenon is again attributable to the viscoelastic characteristics of the FPY pressure sensor. The low magnitude, high-frequency pressure generated during aortic valve closure and its subsequent hemodynamic rebound is attenuated by the viscous resistance of the sensor. Despite this minor attenuation, overall, the FPY pressure sensor has a good linear response and reliably detects the main dynamic features of the ABP waveform. Temporal analysis shows that all main features of the ABP waveform (systolic peak, dicrotic notch, and dicrotic peak) detected by the FPY pressure sensor have accurate timing aligned with the measurement from the OptoForce sensor. Furthermore, the attenuation effect that consistently affects the diastolic peak can be compensated in signal reconstruction using the inverse transfer function.

\section{Conclusions}
In this study, we developed an FPY pressure sensor for continuous non-invasive ABP waveform measurement. The proposed sensor offers the advantage of low-cost materials, a simple fabrication process, and a compact physical design. The evaluation of the FPY pressure sensor in detecting ABP waveforms was performed through two scenarios, using a hardware simulator and direct measurement at an arterial site. Measurements using a simulator were intended to assess the sensor's response to a consistent waveform, while measurements at an arterial site aimed to detect ABP waveforms corresponding to the cardiovascular cycle. The results from the simulator measurement demonstrate that the sensor has excellent repeatability, reproducibility, and accuracy. The measurements at the arterial site also showed high accuracy, as shown by high correlation and temporal synchronization of the ABP waveform's main features with the OptoForce sensor measurements. This is further supported by a low RMSE value, indicating a minor actual difference between the two measurements. According to our results, the FPY pressure sensor presents a novel method for monitoring ABP waveforms. With further development, this sensor holds substantial potential for use as a clinical diagnostic device to assess cardiovascular health and problems. Furthermore, as it is a cost-efficient, small, and comfortable sensor, it can be a component in smart devices (e.g. smartwatches and smart wristbands), enhancing their physiological signal acquisition capabilities and therefore supporting a healthy lifestyle and early diagnosis of cardiovascular disease. To address these requirements, our future work should focus on developing an inverse transfer function for signal reconstruction, especially to compensate for the damping effects caused by the sensor's viscoelastic characteristics, as well as establishing an appropriate calibration method for acquiring absolute blood pressure values.

\vspace{6pt} 

\authorcontributions{Conceptualization, R.M., S.F. and G.C.; data curation, R.M.; formal analysis, R.M. and S.F.; funding acquisition, G.C.; investigation, R.M.; methodology, R.M.; project administration, R.M., S.F. and G.C.; resources, R.M., A.R., S.F. and G.C.; software, R.M.; supervision, A.R., S.F. and G.C.; validation, R.M. and S.F.; visualization, R.M.; writing - original draft, R.M.; writing - review \& editing, S.F. and G.C. All authors have read and agreed to the published version of the manuscript.}

\funding{This research received no external funding.}

\institutionalreview{Ethical review and approval were waived for this study due to the research exclusively involving the author as the single subject.}

\informedconsent{Informed consent was obtained from all subjects involved in the study.}

\dataavailability{The original contributions presented in this study are included in the article. Further inquiries can be directed to the corresponding author.}

\acknowledgments{The authors would like to thank Attila Répai for his technical assistance and expert guidance during the experimental testing using his developed simulator.}

\conflictsofinterest{The authors declare no conflicts of interest.} 

\abbreviations{Abbreviations}{%
The following abbreviations are used in this manuscript:\\

\noindent 
\begin{tabular}{@{}ll}
ABP   & Arterial blood pressure\\
CNIBP & Continuous non-invasive blood pressure\\
FPY   & Flexible piezoresistive yarn\\
LED   & Light-emitting diode\\
PDMS  & Polydimethylsiloxane\\
PPG   & Photoplethysmography\\
RMSE  & Root mean square error\\
ROI   & Region of interest\\
RPM   & Revolutions per minute\\
TPU   & Thermoplastic polyurethane\\
\end{tabular}
}

\reftitle{References}

\PublishersNote{}
\end{document}